\documentclass{acm_proc_article-sp}
\usepackage{parskip}
\usepackage{url}
\usepackage{float}
\usepackage{multirow}
\usepackage{color,xcolor,colortbl}

\definecolor{Gray}{gray}{0.85}

\begin{document}

\title{On predictability of rare events leveraging social media: \\ a machine learning perspective}

\numberofauthors{3} 

\author{
% 1st. author
\alignauthor
Lei Le 	\\
       \affaddr{School of Informatics and Computing, Indiana University }\\
       \affaddr{Bloomington, Indiana 47408}\\
       \email{leile@indiana.edu}
% 2nd. author
\alignauthor
Emilio Ferrara\titlenote{Corresponding author.} \\
       \affaddr{School of Informatics and Computing, Indiana University }\\
       \affaddr{Bloomington, Indiana 47408}\\
       \email{ferrarae@indiana.edu}
% 3rd. author
\alignauthor 
Alessandro Flammini \\
       \affaddr{School of Informatics and Computing, Indiana University }\\
       \affaddr{Bloomington, Indiana 47408}\\
       \email{aflammin@indiana.edu}
%\and  % use '\and' if you need 'another row' of author names
}

\date{February 2015}
% Just remember to make sure that the TOTAL number of authors
% is the number that will appear on the first page PLUS the
% number that will appear in the \additionalauthors section.

\maketitle
\begin{abstract}
Information extracted from social media streams has been leveraged to forecast the outcome of a large number of real-world events, from political elections to stock market fluctuations. An increasing amount of studies demonstrates how the analysis of social media conversations provides cheap access to the wisdom of the crowd. However, extents and contexts in which such forecasting power can be effectively leveraged are still unverified at least in a systematic way. It is also unclear how social-media-based predictions compare to those based on alternative information sources.
To address these issues, here we develop a machine learning framework that leverages social media streams to automatically identify and predict the outcomes of soccer matches.
We focus in particular on matches in which at least one of the possible outcomes is deemed as highly unlikely by professional bookmakers. 
We argue that sport events offer a systematic approach for testing the predictive power of social media conversations, and allow to compare such power against the rigorous baselines set by external sources.
Despite such strict baselines, our framework yields above 8\% marginal profit when used to inform simple betting strategies.
The system is based on real-time sentiment analysis and exploits data collected immediately before the game start, allowing for bets informed by its predictions.
We first discuss the rationale behind our approach, then describe the learning framework, its prediction performance and the return it provides as compared to a set of betting strategies. To test our framework we use both historical Twitter data from the 2014 FIFA World Cup games (10\% sample), and real-time Twitter data (full stream) collected by monitoring the conversations about all soccer matches of the four major European tournaments (FA Premier League, Serie A, La Liga, and Bundesliga), and the 2014 UEFA Champions League,  during the period between October, 25th 2014 and November, 26th 2014.

\end{abstract}

% A category with the (minimum) three required fields
%\category{I.5}{Pattern Recognition}{Miscellaneous}
%\category{J.4}{Social and Behavioral Sciences}{Sociology}
%A category including the fourth, optional field follows...
%\category{D.2.8}{Software Engineering}{Metrics}[complexity measures, performance measures]

%\terms{Experiment, Measurement}

%\keywords{Soccer, Prediction, Twitter, Social Media} % NOT required for Proceedings

\section{Introduction}
A large number of case studies have proved that social media like Twitter can be effective sources of information to understand real-world phenomena and to anticipate the outcomes of events that are yet to happen, like political elections~\cite{digrazia2013more,schoen2013power} and talent shows~\cite{ciulla2012beating}, movies box-office performance~\cite{asur2010predicting,oghina2012predicting}, and stock-market fluctuations~\cite{zhang2011predicting,bollen2011twitter}. Even discounting the fact that successful case studies don't tell much about failures, the effectiveness of social media as information source to predict real events may not be surprising: they offer a window on the collective wisdom of a potentially very large crowd of users that can be harvested at the expense of a relatively small technological investment. On the other hand, a number of potential issues may affect such effectiveness: beyond all sorts of biases in the population of users whose tweets are collected, in virtually all cases the opinion of users can not be directly polled to answer the questions at hand. In some cases, there is arguably a strong correlation between the signal collected and the event to be predicted. The Twitter traffic volume about a movie and the revenue it later generates in the opening week, or the valence of political discussions and the outcome of an election are example of such cases. In others, such correlation is, at least in principle, more tenuous (e.g., the overall mood of Twitter conversations and fluctuations in the stock market). In general, the potential of leveraging information from social media to predict the outcome of real-world events is unclear and certainly has not been systematically studied.

Here we propose that an ideal test bed for addressing this issue is to consider team sport events. They offer several advantages: the number of possible outcomes of sport matches is usually limited, they occur continuously, and there is a lot of potentially useful signal to collect: social media are used by millions of sport fans everyday to discuss about their favorite sports, the teams they cheer for and their performance, and the expectations for future games.
Another non trivial advantage is that prediction based on social media wisdom can be systematically compared with that implicitly reflected in the odds fixed by bookmakers. Betting odds in fact represent the opinion of experienced professionals. Presumably they also take into account the wisdom of the betting crowd, as quotes are continuously re-adjusted to reflect the influx of incoming bets, which in turn can be regarded as proxies of the bettors opinion.

In this paper we discuss the design, implementation, and validation of a machine learning framework to predict the occurrence of very unlikely (in terms of their betting odds) outcomes in soccer games by leveraging the mood of Twitter conversations relative to such games. The choice of soccer was made because it offers a larger Twitter traffic with respect to other sports. Soccer is the most popular sport in the world\footnote{\url{http://mostpopularsports.net/in-the-world/}} with more than 3 billions fans worldwide accourding to recent estimates\footnote{\url{http://www.topendsports.com/world/lists/popular-sport/fans.htm}}. The official blog of Twitter for example reports that there were 672 million tweets posted related to the 2014 FIFA World Cup tournament, making this the most spoken event online in the history of the platform\footnote{\url{https://blog.twitter.com/2014/seven-worldcup-data-takeaways-so-far}}. 

As mentioned above, here we focus on games that have the potential for an outcome deemed very unlikely by bookmakers.
There are at least two reasons for this choice. On the one hand, these games are those potentially more profitable to bet upon, as one of their results has very high odds. More importantly, they are arguably those for which to ``correctly'' estimate the odds is problematic both for bookmakers and bettors, and therefore they offer a potential for successfully leveraging exogenous signals as that extracted from social media.  

We consider games from six different competitions, including the 2014 FIFA World Cup tournament, and the relative Twitter conversations \cite{kim2015social}. 
We extract separately the average mood in the conversations generated by supporters of both teams for a period of six hours before the beginning of the games, and use its discrete representation to train a machine learning classifier called to discriminate between games whose outcome is the expected result (low-odds), or the unlikely one (high-odds). 
Our results translate in a simple betting strategy that offers above 8\%  margin of profit. We interpret this finding as a consequence of both the presence of ``wisdom of the crowd'' signal in social media conversations, and the difficulty to properly estimate the odds of unlikely events. 

Next we present the methodology employed in this study, the procedure used to select the specific games to which our machine learning framework is applied, we introduce the adopted features and then define our classification task. We also offer some intuition on how the selected features, based on the mood of Twitter conversations from the two teams fans, may provide useful information for prediction purpose. In Section~\ref{sec:framework} we describe in detail the implementation of our machine learning approach and its validation according to standard measures of performance. In Section~\ref{sec:profit} we assess the economic profit yielded by using our framework introducing a simple betting strategy based of the results of our predictions. We finally discuss further details on data collection and related work in sections \ref{sec:data} and \ref{sec:related} respectively. We conclude with a summary of our results and a discussion of their relevance.

\section{Methodology}
We considered games from six different tournaments: \emph{(i)} the 2014 FIFA World Cup tournament, \emph{(ii)} the major four European national tournaments during 2014 (FA Premier League, Serie A, La Liga, and Bundesliga), and \emph{(iii)} the 2014 UEFA Champions League. While for the FIFA World Cup we collect historical data from our Twitter \emph{gardenhose} repository at Indiana University (containing about 10\% of the entire datastream), the conversation about the other events is collected using a real-time monitoring algorithm processing the full Twitter stream. In the following, we will consider two datasets: the FIFA, consisting of the games in the the 2014 FIFA World Cup tournament, and one with the games from all other tournaments. We will refer to this second dataset as ``Live-monitoring''.

We are specifically interested in games that before their starting had a potential outcome deemed as very unlikely. For a generic game $g$ we considered the average odds (the latest available before the start of the match) assigned to that match by multiple bookmakers\footnote{The odds can be found at \url{http://odds.sports.sina.com.cn/liveodds/} They are the average of decimal odds rather than Asian Handicap odds from all accessible betting companies. All odds are the last updated ones before the game.}. We leverage four betting agencies: William Hill, Ladbrokes, Bet 365, and Bwin; these four bookmakers together cover most of the betting market. We define $O^{g}_{\max}$ and $O^{g}_{\min}$ as the maximum and minimum odds assigned to one of the possible outcomes of the game $g$. We also define $O^{g}$ as the the odds of the outcome that finally materializes. Of course $O^{g}$ can coincide with one of $O^{g}_{\max}$ and $O^{g}_{\min}$. To each game we assigned a \emph{potential upset} score $PU(g)$ that measures the relative likelihood of the most likely outcome to the least one

\begin{equation}
PU(g)=\frac{O^{g}_{\max}-1}{O^{g}_{\min}-1}.
\label{eq:potential_upset_score}
\end{equation}

Note that, in betting, larger odds identify less likely outcomes: from Eq. \ref{eq:potential_upset_score} it follows that the higher the upset score for a game, the more unlikely that outcome was according to the bookmakers. Eq. \ref{eq:potential_upset_score} has lower bound at 1 and no theoretical upper bound: the practical upper bound is determined by how disproportionate the game odds are; in our experience the upset score max out around 100. Correctly betting on games turning into unexpected outcomes could generate the largest marginal profits if correctly bet upon.

The subset of games relevant to our prediction task are those whose $PU$ score exceeds a given arbitrary large threshold $\theta$. We considered various values of threshold, and in this study we report the results for $\theta = 5$; consistent findings hold for other values in the range $3\leq\theta \leq5$. We finally define the \emph{upset} score $U(g)$ of a game $g$ as the relative likelihood of the outcome that finally materializes to the most likely

\begin{equation}
U(g)=\frac{O^{g}-1}{O^{g}_{\min}-1}.
\label{eq:upset_score}
\end{equation}

\begin{figure}[!t]
\centering
\includegraphics[width=.8\columnwidth,clip=true,trim=10 10 10 10]{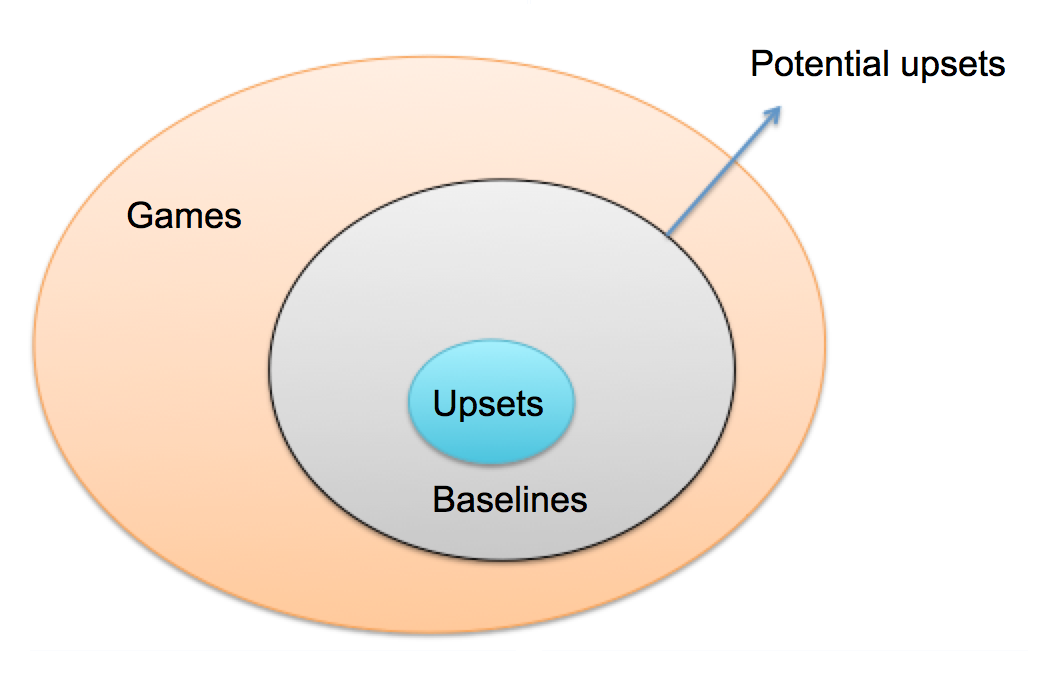}
\caption{Upsets, potential upsets and baselines. }%``Upsets'' are included in ``Potential upsets'' and ``Baselines'' are the set difference between ``Potential upsets'' and ``Upsets''}
\label{fig:stylized}
\end{figure}

U(g) can be as small as 1 when the most likely result (minimum odds) materializes and as big as $PU(g)$ if the least likely result occurs. For illustrative purposes, in Table \ref{tab:2014fifa} we show the list of all FIFA World Cup games with an outcome different from the most likely ($U(g)>1$). This happened for 31 games out of 64 played during the 2014 tournament. In the following, we will refer to games with $PU(g) > \theta$ as \emph{potential upsets} and to games with $U(g) > \theta$ as \emph{upsets}. Given the definition above, the latter constitute a subset of the former, as depicted in Fig.~\ref{fig:stylized}.  We will refer to games that are potential upsets but not upsets as baseline games. 

From Table \ref{tab:2014fifa}, the reader knowledgeable of soccer will immediately see that some games with very unlikely scores (for example Brazil 1:7 Germany) are attached with low upset scores: this because our framework ignores goal differences and considers only for the overall outcome of a match. On the other hand, largely unexpected defeats like Uruguay 1:3 Costa Rica and Italy 0:1 Costa Rica, or ties like Germany 2:2 Ghana or Brazil 0:0 Mexico, yield  large upset scores.

Note that a potential upset game can be an upset without necessarily resulting in $U(g)=PU(g)$. Consider the following example game between team $A$ and $B$ whose odds are (2,7,11) on the victory of team A, a draw, and the victory of team B, respectively. The game is a potential upset according to our threshold $\theta = 5$, because $PU(g)=(11-1)/(2-1) = 10 > \theta =5$. Suppose that the final outcome is a draw.
The game is an upset because $U(g) = (7-1)/(2-1) = 6  > \theta =5$, but $U(g) < PU(g)$. Interestingly, although this is a possibility, we never observed any such case in our datasets (see Tables~\ref{tab:fifa2014training} and~\ref{tab:league2014training}).

Our classification tasks will consist in discriminating games that turn out to be \emph{upsets} among all \emph{potential upsets} using features extracted from Twitter conversations relative to such games. We discuss the details about the data collection Section \ref{sec:data}. Before turning to a detailed description of our framework and of the features it employs, in the next section we provide some support to the idea that Twitter conversations may reflect
important information about a game, which in turn can be  leveraged to predict its outcome.

\begin{table}[!t]
\centering \small
\caption{Upset scores for the 2014 FIFA World Cup upset games. \emph{(a.e.t.}: result after extra time)\bigskip}
\begin{tabular}{|c|c|}
\hline
Game & U(g)\\
%\begin{tabular}{|c|c|c|}
%\hline
%game&upset score&UTC time\\
\hline\hline
Uruguay 1:3 Costa Rica&18.04\\
\hline
Germany 2:1(a.e.t.) Algeria&15.03\\
\hline
Germany 2:2 Ghana&14.06\\
\hline
Brazil 0:0 Mexico&12.87\\
\hline
Italy 0:1 Costa Rica&9.69\\
\hline
Spain 0:2 Chile&7.96\\
\hline
Brazil 3:2(a.e.t.) Chile&6.46\\
\hline
Netherlands 4:3(a.e.t.) Costa Rica&6.22\\
\hline
Argentina 1:0(a.e.t.) Switzerland&5.72\\
\hline
Ecuador 0:0 France&5.09\\
\hline
Spain 1:5 Netherlands&4.86\\
\hline
USA 2:2 Portugal&4.08\\
\hline
Costa Rica 0:0 England&3.79\\
\hline
Nigeria 1:0 Bosnia Herzegovina&3.79\\
\hline
Russia 1:1 Korea Republic&3.17\\
\hline
Greece 2:1 C\^{o}te d'Ivoire&3.04\\
\hline
Iran 0:0 Nigeria&2.93\\
\hline
Uruguay 2:1 England&2.86\\
\hline
Algeria 1:1 Russia&2.61\\
\hline
Belgium 2:1(a.e.t) USA&2.39\\
\hline
Brazil 0:3 Netherlands&1.75\\
\hline
Japan 0:0 Greece&1.71\\
\hline
Germany 1:0(a.e.t.) Argentina&1.70\\
\hline
Netherlands 2:4(a.e.t.) Argentina&1.46\\
\hline
England 1:2 Italy&1.46\\
\hline
Ghana 1:2 USA&1.40\\
\hline
Korea Republic 2:4 Algeria&1.31\\
\hline
Costa Rica 5:3(a.e.t.) Greece&1.26\\
\hline
Italy 0:1 Uruguay&1.15\\
\hline
Brazil 1:7 Germany&1.14\\
\hline
Netherlands 2:0 Chile&1.01\\
\hline
\end{tabular}
\label{tab:2014fifa}

\end{table}

\bigskip

\section{Interpreting the game signals}
Excluding extra time and penalties, a soccer game usually lasts less than 120 minutes with two 45-minute halves, a 15-minute halftime break, and several minutes of injury time. In this section, we seek to understand how well Twitter reflects the events occurring during a soccer game. For this in-depth analysis, we focus on the 2014 FIFA World Cup matches, and for simplicity we analyze the Twitter conversation occurring during the 120 minutes representing the effective duration of each game, at the minutes resolution. 

\subsection{Events and Response}
\begin{figure*}[!t]
\centering
\includegraphics[width=1.8\columnwidth,clip=true,trim=10 0 10 10]{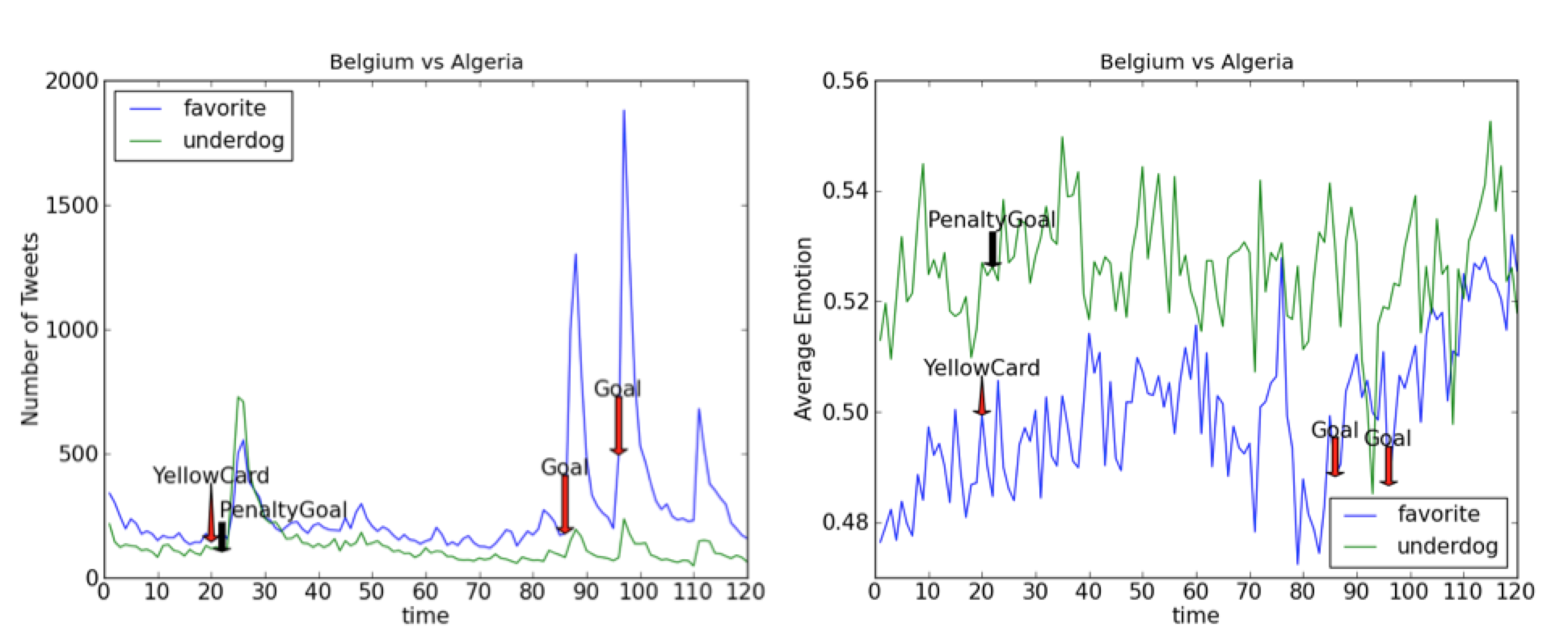}
\caption{\textbf{Events and response during games: volume (left) and average emotion (right) of  tweets.}}
\label{fig:response}
\end{figure*}

We start trying understanding how users respond to important events during a soccer game. 
We only considered the events defined in the official match report provided by FIFA: ``Goal scored,'' ``Penalty scored,'' ``Yellow Card,'' and ``Red Card''. By manually analyzing five upsets and five baseline games, we noticed that in both cases, the number of tweets spikes for a few minutes after these events occur. ``Penalty scored'' is somehow an exception because the number of tweets spikes before this type of events happens, as expected since ``Penalty scored'' occurs shortly after another unrecorded event, namely ``Penalty decision''.
Fig. \ref{fig:response} shows one example of such collective reactions, for the game ``Belgium vs Algeria''. Clear spikes of traffic are annotated with in-game events, which also trigger big fluctuations in the collective sentiment scores (the technical details about sentiment analysis are in Section \ref{sub:sa}): the underdog fans' average sentiment is consistently much lower than the favorites' one, and drops drastically twice as an immediate consequence of the favorite team scoring. This type of analysis shows how well the Twitter conversation captures in real time the collective mood of the supporters, in support of our high-level idea that social media signals can be used to sense live events, and possibly even predict rare ones.

\subsection{Interaction of Groups}

\begin{figure}[!t]
\centering
\includegraphics[width=\columnwidth]{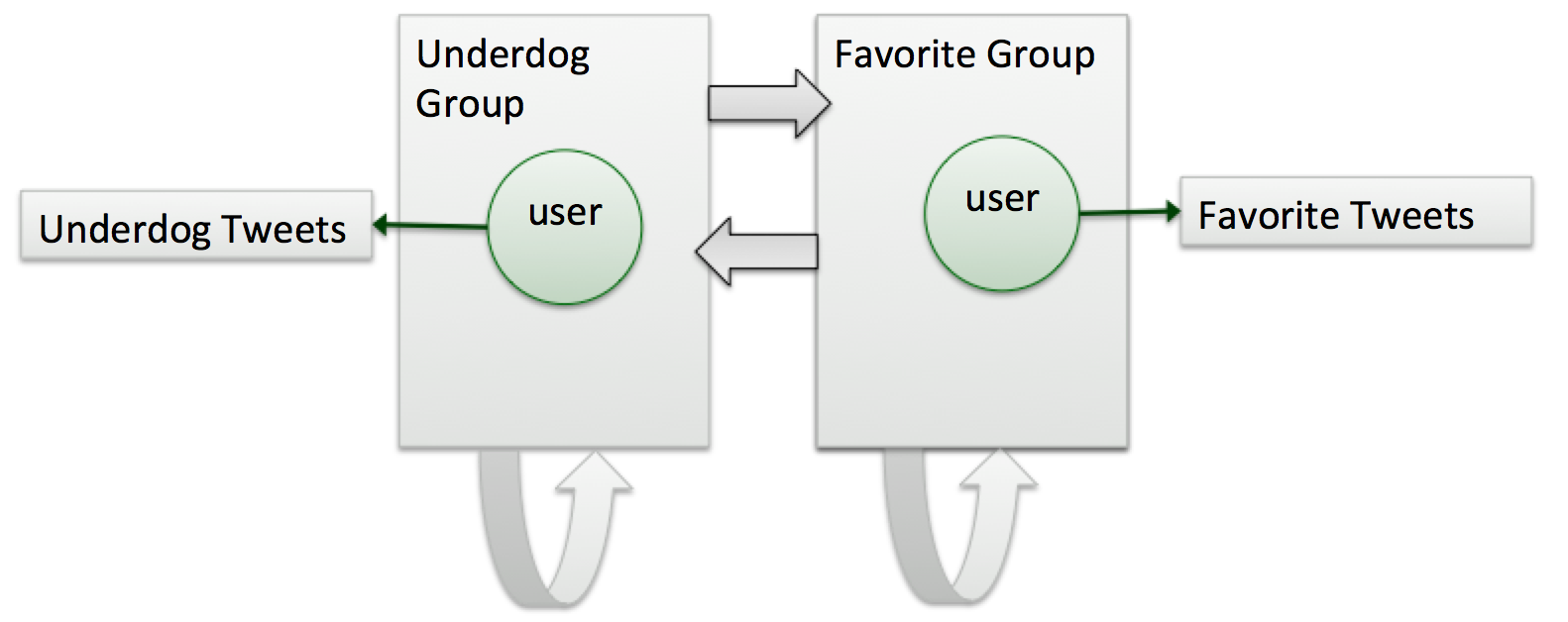}
\caption{Groups and their possible interactions.} 
\label{fig:interaction_schema}
\end{figure}

\begin{figure}[!t]
\centering
\includegraphics[width=\columnwidth,clip=true,trim=10 10 10 10]{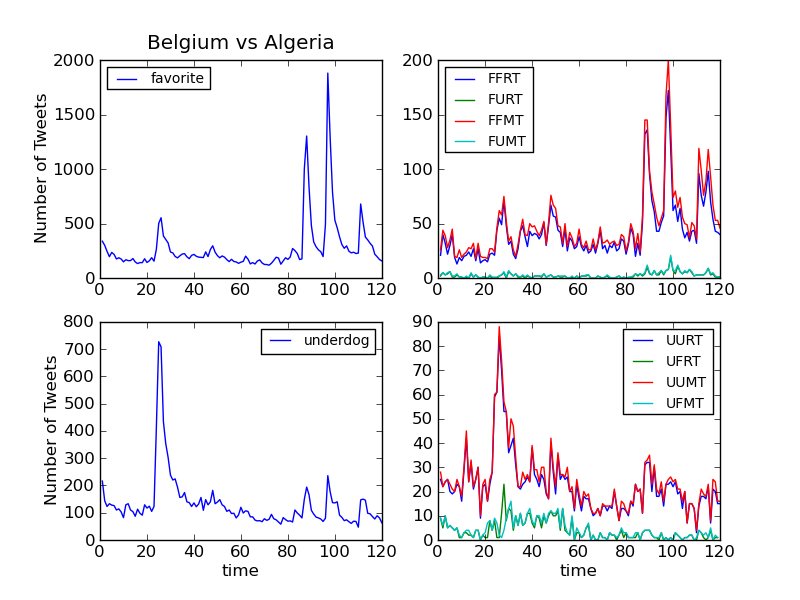}
\caption{Interactions between and within groups during games. %\emph{RT} are retweets, \emph{MT} are mentions, \emph{U} is the group of underdog fans and \emph{F} the group of favorite fans.
\emph{FFRT}/\emph{FFMT}: retweets and mentions within Favorite fans; \emph{FURT}/\emph{FUMT}: retweets and mentions from Favorite fans to Underdog fans. Viceversa, \emph{UURT}/\emph{UUMT}: retweets and mentions within Underdog fans; \emph{UFRT}/\emph{UFMT}: retweets and mentions from Underdog fans to Favorite ones.}
\label{fig:interactions}
\end{figure}

We divided the users tweeting during a game into two groups. One group contains the fans of the favorite team (users who only tweet using the name/abbreviation of the favorites) while the other contains only supporters of the underdog team (see Section \ref{sec:data} for the details about the data collection). We assume that these two groups represent the two factions of supporters. We want to study the interaction dynamics within and between these two groups. The interaction can be in the form of \emph{retweets} or \emph{mentions} to users within the same group or from the other group. Fig. \ref{fig:interaction_schema} schematizes this dynamics. Our analysis shows  that the volume of interactions within groups greatly outnumbers that between groups: Fig. \ref{fig:interactions} illustrates this for the example game ``Belgium vs Algeria''. 
%This suggests the existence of a collective group identity, as hypothesized by recent studies in social psychology. 
%We also studied the sentiment score of these interaction tweets and performed the U-test again. Although the sentiment score of tweets within both groups was not significantly different between upsets and baselines, it was significantly different for tweets both from favorite group to underdog group and the opposite direction. It seems the occurrence of upsets will affect users' opinions when interacting with members from other groups. 

%\subsection{Live-monitoring soccer games data}
%During the period between October, 25th 2014 and November, 26th 2014 we monitored the odds of all games for the four European national tournaments and the UEFA Champions League. Among all games, our system selected 55 potential upsets with a profitability of at least 5:1 ($\theta=5$) in case the most unexpected result had occurred.
%We collected tweets about these 55 games using Twitter Streaming API for 55 soccer games. 
%Among these games, 16 of them turned into upsets (the underdog either won or tied) while 39 were baseline games (the favorite won). As for the 2014 FIFA World Cup games, we selected these game tweets based on hashtags containing teams' abbreviations. However, the adoption of the Streaming API ensured also that we collected (nearly) 100 percent of the relevant tweets, rather than a sample. Some games, however, do not have enough tweets for analysis. Our tuning led us to filter out those games that did not collect at least 40 tweets per team per hour. This post-processing yielded a dataset 31 games, of which 9 turned into upsets and 22 were baselines. The final dataset of league matches contains 1278485 tweets, including 521776 retweets and 31281 replies, produced by 1009034 unique users.

\section{The prediction framework} \label{sec:framework}
Our framework relies on the intuition that fans' discussion preceding a soccer game might convey useful information to predict the outcome of the game.
%To this aim, we can build timeseries representing specific features of the game conversation measured over time: 
Here in particular we seek to exploit the temporal evolution of the sentiment extracted from the Twitter conversations of the opposite set of fans to predict the outcome of potential upset games. 
We argue that sentiment analysis may help uncovering the hopes and therefore the collective opinion about the outcome of the game. The basic assumption is supported by recent social and behavioral psychology studies on social attention~\cite{mason2007situating,shteynberg2013power,shteynberg2014feeling}: in a situation of perceived advantage, the fans of the favorite team will collectively express more positive emotions and feelings than the fans of the opposing team. Our working assumption is, therefore, that games where such gap in positive emotions is not observed before the game starts will consistently turn into upsets. In the following we describe our effort to test such assumption.

\bigskip

\subsection{Testing the significance of sentiment gap}
We computed the sentiment score for each tweet produced either by the favorite or the underdog supporter in a 6 hours time period preceding the beginning of the game: tweets sentiment scores range in the interval $[0,1]$ (see Section \ref{sec:data} for details).
For each game, we retrieved the Twitter conversation occurred during the 6 hours before the start, and we broke this period into 12 windows (each representing 30 minutes) and computed the distribution of sentiment score in each window for tweets from the favorite and the underdog supporters, separately.\footnote{We explored alternatives, including sliding windows with partial overlap and different window lengths. The configuration reported here yields the best performance. We also exclude match-related tweets (those mentioning both teams) to avoid deciding how to attribute that sentiment the teams.
}

We finally represented each game with a single vector $P(g)=\{p_1, \dots, p_{12}\}$, where each component 
%$p_i=U(t_i^f, t_i^u)$ 
is the p-value of the Mann-Whitney U-test between the distribution of sentiment expressed toward the favorite and the underdog team during the $i^th$ time window.

\begin{figure}[!t]
\centering
\includegraphics[width=\columnwidth,clip=true,trim=0 20 0 20]{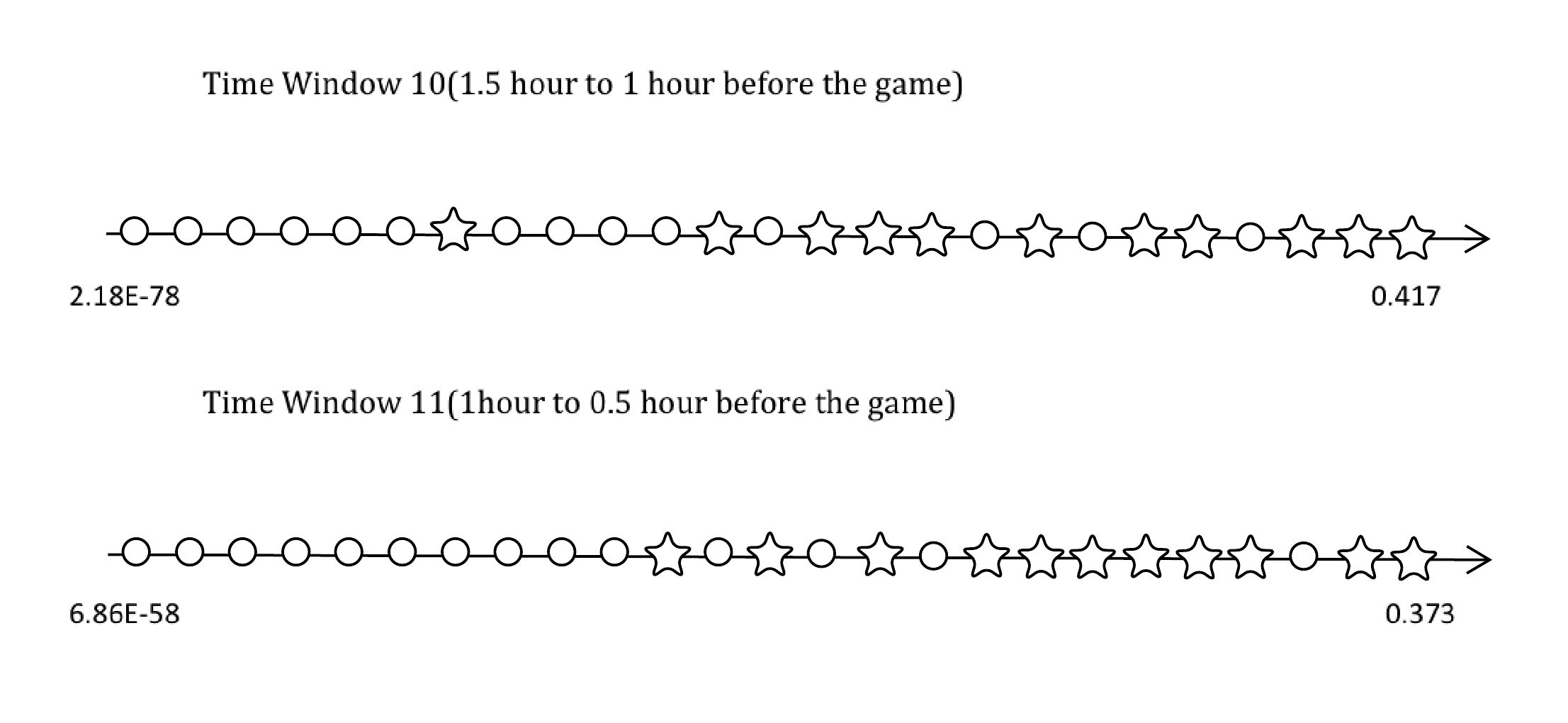}
\caption{Predictions based on sentiment score gap. Each star/circle denotes a game prediction. Stars denote games predicted as upsets, circles are games predicted as baseline. The axis denotes the p-values. }
\label{fig:classifier}
\end{figure}

Tables \ref{tab:utest1} and \ref{tab:utest2} show the results under the significance level of p<0.0001 for the two datasets (FIFA and Live-monitoring respectively). 
%Ideally, if our two hypotheses are true, we should observe that no upset games pass the U-test, while all baseline games pass it. 
%What we can observe in reality is that, 
When one considers early time windows our hypothesis fails, as most of the games don't pass the U-tests, regardless of their final result. 
However, when one considers later time windows (e.g., time windows 10 and 11, which is 90 minutes to 30 minutes before the games start), the majority of baseline games pass the U-test, while only a small fraction of upsets do (see Fig. \ref{fig:classifier}). This suggests that a significant difference in sentiment distribution between the two factions of fans is discriminative in identifying games that turn into upsets.

%This is good news: we can use the late Twitter signals before the game starts, taking advantage of the difference in sentiment collectively express by the fans of the two teams, to make a prediction. 
%To this purpose, time window 10 and 11 are valuable to distinguish upset games from baseline games. 

Most of the usable sentiment signal is conveyed between 90 minutes and 30 minutes before the games start.
For readers knowledgeable of soccer, such information won't be surprising: line-ups are usually announced about 90 minutes before the games. Releasing news on line-ups and other factors of the game, such as last-minute injuries, the weather, etc., may influence the opinions of the fans about the outcome of the game. 

\begin{table}[!t]
\centering \small 
\caption{U-test on sentiment scores (p<0.0001) on the FIFA World Cup dataset. Ideally, upsets should pass no tests, and baselines should pass all test\bigskip}
\begin{tabular}{|c|c|c|}
\hline
Window & Upset (pass/total) & Baseline (pass/total)\\
\hline\hline
 1 & 3/10 & 5/15\\
\hline
 2 & 3/10 & 3/15 \\
\hline
 3 & 2/10 & 2/15 \\
\hline
 4 & 3/10 & 3/15 \\
\hline
 5 & 3/10 & 3/15 \\
\hline
 6 & 2/10 & 4/15 \\
\hline
 7 & 3/10 & 3/15 \\
\hline
 8 & 2/10 & 3/15 \\
\hline
 9 & 3/10 & 3/15 \\
\hline\rowcolor{Gray}
 10 & 3/10 & 10/15 \\
\hline\rowcolor{Gray}
 11 & 3/10 & 11/15 \\
\hline
 12 & 7/10 & 8/15 \\
\hline
\end{tabular}
\label{tab:utest1}
\end{table}

\begin{table}[!t]
\centering
\caption{U-test on sentiment scores (p<0.0001) on the Live-monitoring dataset. Ideally, upsets should pass no tests, and baselines should pass all test\bigskip}
\begin{tabular}{|c|c|c|}
\hline
Window & Upset (pass/total) & Baseline (pass/total)\\
\hline\hline
 1 & 1/9 & 0/22\\
\hline
2 & 5/9 & 3/22 \\
\hline
 3 & 0/9 & 2/22 \\
\hline
4 & 2/9 & 7/22 \\
\hline
 5 & 0/9 & 0/22 \\
\hline
 6 & 3/9 & 11/22 \\
\hline
 7 & 0/9 & 2/22 \\
\hline
8 & 4/9 & 10/22 \\
\hline
 9 & 1/9 & 2/22 \\
\hline\rowcolor{Gray}
10 & 2/9 & 14/22 \\
\hline\rowcolor{Gray}
 11 & 2/9 & 11/22 \\
\hline
 12 & 3/9 & 12/22 \\
\hline
\end{tabular}
\label{tab:utest2}
\end{table}

\begin{table}[!t]
\centering \small
\caption{The 2014 FIFA World Cup training set ($\theta=5$): upset  and baseline games\bigskip}
\begin{tabular}{|@{}c@{}|c|c|c|c|}
\hline
Game & U(g) & PUS(g) & Class \\
\hline\hline
Uruguay 1:3 Costa Rica & 18.05 & 18.05 &upset\\
\hline
Germany 2:1(a.e.t) Algeria & 15.03 & 30.2 &upset\\
\hline
Germany 2:2 Ghana & 14.06 & 24.47 &upset\\
\hline
Brazil 0:0 Mexico & 12.88  & 12.88 &upset\\
\hline
Italy 0:1 Costa Rica & 9.69  & 9.69 &upset\\
\hline
Spain 0:2 Chile & 7.96  & 7.96 &upset\\
\hline
Brazil 3:2(a.e.t) Chile & 6.46  & 9.98 &upset\\
\hline
Netherlands 4:3(a.e.t) Costa Rica & 6.22  & 12.04 &upset\\
\hline
Argentina 1:0(a.e.t) Switzerland & 5.72  & 9.98 &upset\\
\hline
Ecuador 0:0 France & 5.1  & 7.25 &upset\\
\hline
Cameroon 1:4 Brazil & 1.0 & 166.25 & baseline\\
\hline
Argentina 1:0 Iran & 1.0  & 145.69 & baseline\\
\hline
Australia 2:3 Netherlands & 1.0  & 53.59 & baseline\\
\hline
France 3:0 Honduras & 1.0  & 41.81 & baseline\\
\hline
Brazil 3:1 Croatia & 1.0  & 35.5 & baseline\\
\hline
Argentina 2:1 Bosnia H. & 1.0 & 29.43 & baseline\\
\hline
Belgium 2:1 Algeria & 1.0 & 25.97 & baseline\\
\hline
Nigeria 2:3 Argentina & 1.0  & 19.43 & baseline\\
\hline
Chile 2:1 Australia & 1.0  & 18.41 & baseline\\
\hline
France 2:0 Nigeria & 1.0  & 17.71 & baseline\\
\hline
Australia 0:3 Spain & 1.0  & 17.39 & baseline\\
\hline
Honduras 0:3 Switzerland & 1.0  & 12.7 & baseline\\
\hline
USA 0:1 Germany & 1.0  & 10.52 & baseline\\
\hline
Honduras 1:2 Ecuador & 1.0  & 7.85 & baseline\\
\hline
Cameroon 0:4 Croatia & 1.0  & 6.89 & baseline\\
\hline
\end{tabular}
\label{tab:fifa2014training}
\end{table}

\begin{table}[!t]
\centering \small
\caption{The European leagues games training set ($\theta=5$): upset  and baseline games\bigskip}
\begin{tabular}{|@{}c@{}|c|c|c|c|}
\hline
Game & U(g) & PUS & Class\\
\hline\hline
Dortmund 0:1 Hannover 96  & 91.66 & 91.66 & upset\\
\hline
Liverpool 0:0 Hull City  & 14.44  & 24.07 & upset\\
\hline
West Ham  2:1 Manchester City  & 14.85  & 14.85 & upset\\
\hline
Tottenham  1:2 Newcastle Utd  & 9.65  & 9.65 & upset\\
\hline
Milan 0:2 Parlemo  & 15.27  & 15.27 & upset\\
\hline
Arsenal 3:3 Anderlecht  & 20.47  & 39.05 & upset\\
\hline
Manchester City 1:2 CSKA   & 44.73  & 44.73 & upset\\
\hline
QP Rangers 2:2 Manchester City  & 9.44  & 14.44 & upset\\
\hline
Real Sociedad 2:1 Atletico Madrid  & 6.63  & 6.63 & upset\\
\hline
Southampton 1:0 Stoke City  & 1.0 & 12.68 & baseline\\
\hline
Sunderland 0:2 Arsenal  & 1.0 & 6.60 & baseline\\
\hline
Cesena 0:1 Inter  & 1.0 & 11.70 & baseline\\
\hline
Juventus 2:0 Palermo  & 1.0 & 100.00 & baseline\\
\hline
Napoli 6:2 H. Verona  & 1.0 & 25.18 & baseline\\
\hline
Arsenal 3:0 Burnley  & 1.0 & 78.57 & baseline\\
\hline
Bayern Munich 2:1 Dortmund  & 1.0 & 10.97 & baseline\\
\hline
Empoli 0:2 Juventus  & 1.0 & 27.85 & baseline\\
\hline
Granada 0:4 Real Madrid  & 1.0 & 84.61 & baseline\\
\hline
Dortmund 4:1 Galatasaray  & 1.0 & 65.00 & baseline\\
\hline
Juventus 3:2 Olympiacos  & 1.0 & 42.85 & baseline\\
\hline
Malmo 0:2 Atletico Madrid  & 1.0 & 35.87 & baseline\\
\hline
Real Madrid 1:0 Liverpool  & 1.0 & 66.07 & baseline\\
\hline
Ajax 0:2 Barcelona  & 1.0 & 26.80 & baseline\\
\hline
Bayern Munich 2:0 Roma  & 1.0 & 67.85 & baseline\\
\hline
PSG 1:0 Apoel  & 1.0 & 75.00 & baseline\\
\hline
Manchester Utd 1:0 Crystal Palace  & 1.0 & 42.50 & baseline\\
\hline
Roma 3:0 Torino  & 1.0 & 18.82 & baseline\\
\hline
Dortmund 1:0 Borussia M.  & 1.0 & 7.55 & baseline\\
\hline
Wolfsburg 2:0 Hamburg  & 1.0 & 9.33 & baseline\\
\hline
Juventus 7:0 Parma  & 1.0 & 92.30 & baseline\\
\hline
PSG 2:0 Olympique  Marseille  & 1.0 & 8.87 & baseline\\
\hline
\end{tabular}
\label{tab:league2014training}
\end{table}

\subsection{Prediction}
As anticipated above, the primary goal of this paper is to describe a machine learning  framework that, among all potential upset games, discriminate those that \emph{actually} turn into an upset. In the datasets considered here, based on the odds we collected, any result other than the victory of the favorite team will make the game an upset; therefore, our classification task can be rephrased as discriminating between the victory of the favorite and either a draw or the victory of the underdog. 

We considered different classification approaches, all based on the feature vector $P(g)$ defined above.
%a single vector $\{p_1, \dots, p_{12}\}$, where each component $p_i=U(t_i^f, t_i^u)$ is the p-value of the Mann-Whitney U-test between the distribution of sentiment expressed toward the favorite team ($t_i^f$) and that for the underdog team ($t_i^u$) during the 30min-long time window $i$. 

We explored the performance of most classifiers available in the Python library scikit-learn~\cite{scikit-learn}: the best performance is provided by Gaussian Naive Bayes. Note that our goal here was not that of finding the best classifier or the best parameter tuning, but to illustrate the feasibility of our method: more advanced machine learning techniques, such as deep learning, might yield even better performance.
We use the two datasets (FIFA and Live-monitoring) to train our classifier and then perform a stratified three-fold cross validation to evaluate its performance, which are shown in Tables \ref{tab:fifa_performance} and \ref{tab:live_performance}. Data about the World Cup were collected from the Twitter gardenhose (10\% sample), while those in the ``Live-monitoring'' set from the Twitter Streaming API (full stream).
We decided to keep these two sets separate as they exhibit sensibly different volumes of tweets, due to the magnitude of the events and the sampling rate of the Twitter streams.

Let us discuss these two cases separately. Table \ref{tab:fifa_performance} illustrates the prediction performance with the 25 potential upsets that constitute our FIFA World dataset.
%, using the signal from timewindow $i=11$, which means Twitter conversation generated only between 1 hour and 30 minutes before the beginning of the game. This window yields the best prediction performance and represents the best trade-off between amount of data used and accuracy of the prediction. 
Our classifier in this scenario achieves an accuracy near to 79\% and a score in terms of AUROC near to 73\%. The results based on the Twitter gardenhose are promising, but we expect to be able to do even better with live-monitoring the games using the full Twitter stream. Table \ref{tab:live_performance} shows the performance for the 31 potential upsets identified during the period between October, 25th 2014 and November, 26th 2014 in the four major European national tournaments plus the UEFA Champions League. 

%Examples of classifications using two time windows ($i=10$ and $i=11$ respectively) are shown in Fig. \ref{fig:classifier}.

In the case of live-monitoring games, we can improve our prediction performance scoring an accuracy of 83.63\% and an AUROC of 78.87\%. These results clearly suggest that our framework can be potentially used for early prediction of the games. 
%However, before assessing what is the marginal profit that we can obtain by betting following the predictions of our system, we want to exclude the possibility that the results obtained by the classifier are due to chance. To this aim, 
As a proof of consistency, given the relatively small set of potential upset games, we constructed two randomized versions of the datasets in which we randomly reshuffle the class labels of each game (upset or baseline game) across all games. This process yields a yardstick in which sentiment is disentangled from the actual game results. As Table \ref{tab:baseline} shows, both Accuracy and AUROC in such random model classification exhibit scores near 50\%, confirming the presence of predictive signal in our game representation.

%
%
%The performance is comparable to our previous findings. We also checked the U-test on the sentiment score of favorites and underdogs for these 31 games. The same distinction appears from time window 10. However, during these matches, time window 12 shows a difference between upsets and baselines. Since most of such matches are not as popular as World Cup games, and considering potential upsets are not high profile prior to the game by nature, it's unlikely those games are not as attractive to fake fans. In fact, we did not observe the surging number of tweets during the last 30 minutes before the game. Therefore, this observation can somewhat support our argument that the noise during the last time window disturb the distinction between upsets and baselines. 

Based on all results and observations above, we concluded we can make highly profitable predictions on potential upset games based solely upon the difference of sentiment expressed by the fans of the two teams prior to the match. Specifically, in the range between 90 to 30 minutes before the games start, the difference of sentiment scores between favorites and underdogs is usually significant for baseline games and not significant for upsets. We leverage this prediction framework next, to determine what margin of profit we can achieve betting on potential upsets, as compared to other betting strategies not informed by social media data.

\begin{table}[!t]
\centering
\caption{Classification performance of historical games (2014 FIFA World Cup)\bigskip}
\begin{tabular}{|c|c|c|c|c|}
\hline
Accuracy & Precision & Recall & F1-Score & AUROC\\
\hline\hline
0.7898 & 0.8512 & 0.5431 & 0.6631 & 0.7286 \\
\hline
\end{tabular}
\label{tab:fifa_performance}
\end{table}

\begin{table}[!t]
\centering
\caption{Live-monitoring game prediction performance (2014 European tournaments)\bigskip}
\begin{tabular}{|c|c|c|c|c|}
\hline
Accuracy & Precision & Recall & F1-Score & AUROC\\
\hline\hline
0.8363 & 0.5833 & 0.6667 & 0.6190 & 0.7887 \\
\hline
\end{tabular}
\label{tab:live_performance}
\end{table}

\begin{table}[!t]
\centering
\caption{Classification performance on reshuffle model for baseline comparison\bigskip}
\begin{tabular}{|c|c|c|c|c|}
\hline
Accuracy & Precision & Recall & F1-Score & AUROC\\
\hline\hline
0.5576 & 0.45	& 0.3	&	0.3428 & 0.5116 \\
\hline
\end{tabular}
\label{tab:baseline}
\end{table}

%\begin{figure}[!t]
%\centering
%\includegraphics[width=\columnwidth,clip=true,trim=0 20 0 20]{r1.png}
%\caption{Prediction and Sentiment Score. All the circles denote games. If the color is red, the game is predicted as upset; if the color is blue, the game is predicted as baseline. The axis denotes the p-values. }
%\label{fig:classifier}
%\end{figure}

\section{Economic Profit on Predictions} \label{sec:profit}
The ultimate test of the effectiveness of the predictive power of our approach consists in determining whether it can return a profit if used systematically against the odds offered by the bookmakers. Such odds are notoriously hard to beat because: \emph{(i)} they are initially set by  professional soccer experts, \emph{(ii)} they are continuously adjusted to take into account the incoming flow of bets (and therefore they take indirectly into account the wisdom of the bettor crowd), and \emph{(iii)} they incorporate a systematic profit margin for the bookmakers. In other words, given the underlying probability of an event to occur, the profit for a successful bettor is less than that would be entitled to in a fair bet. 

Here, we first estimate the average return of a betting strategy based on our predictions, and then compare it with that achievable with different baseline betting strategies. 

\begin{figure}[!t]
\centering
\includegraphics[width=.75\columnwidth,clip=true,trim= 5 5 5 5]{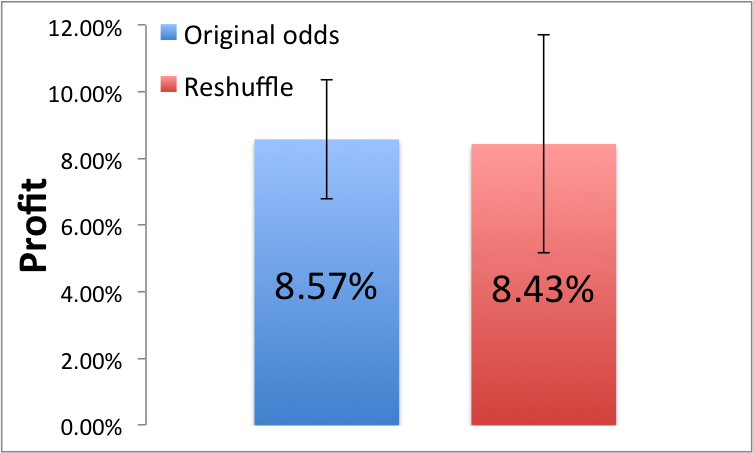}
\caption{Average profits above 8\% yielded by betting according to our predictions.} 
\label{fig:profit1}
\end{figure}

Our two datasets combined contain a total of $N=56$ games (25 potential upsets from the 2014 FIFA World Cup and 31 potential upsets from the live-monitoring European tournaments). We perform 100 rounds of betting. For each round we perform a stratified three-fold cross validation and bet 1 dollar in each of the games in the test set according to the  following simple strategy: if our system predicts that the game will not turn into an upset, we bet the dollar on the favorite team; otherwise, we bet half dollar on the victory of the underdog, and half dollar on the draw. In our datasets, both the latter two results ---if realized--- make the corresponding game an upset and therefore offer a return at least $\sigma = 5$ times larger than the victory of the favorite.
We then compute the marginal profit for the given betting round as
 
\begin{equation}
P=\frac{r-b}{b}
\label{eq:profit}
\end{equation}

where $r$ and $b$ are the total payoff and money bet, respectively. 
Clearly, if $r<b$, Eq. \ref{eq:profit} is negative, which means to incur in a loss rather than a profit ($P<0$). Finally, we compute the average and standard deviation of the marginal profit across all the betting round. The result is represented in the blue bar in Fig. \ref{fig:profit1}.
The average marginal profit of $8.57\%$ is surprisingly high. One possible explanation we wish to exclude is that we consistently classify correctly a single (or few games) with very high return, which would possibly offset and hide a large number of less profitable mis-classifications.
We therefore performed an experiment analogous to the one described above, but where the three odds relative to game are randomly reshuffled across the games.
The average marginal profit is $8.43\%$ and, again, surprisingly high (see red bars in  Fig. \ref{fig:profit1}). This demonstrates that our results are not an artifact of a possibly peculiar odds distribution.

We adopted a stratified three-fold cross validation procedure on the 56 potential upset games and evaluated the results of our predictor on the testing set every time. 
%Since the profit can be sensitive to the odds of the games, we also randomly swapped the odds of some upsets and performed the same strategy as further test. The results are shown in Fig. \ref{fig:profit1}: the three groups of bars identify the three results based on three testing sets. Blue bars stand for the results based on the original odds, while red bars stand for the results based on the reshuffled odds. All the three results using the original odds clearly show that our system yields a marginal profit between 7.89\% and 12.2\% (mean $\mu=9.54\%$ and std. $\sigma=2.32\%$). While the three results based on reshuffled odds exhibit smaller scores (mean $\mu=6.9\%$ and std. $\sigma=2.26\%$) than previous ones, they still provide a great marginal positive profit. To the best of our knowledge, our is the best predictive system on soccer games based on social media streams data as of date. 
This simple strategy that bets equally on all games, regardless on their potential upset score, provides a systematic advantage and marginal profit. We tested more advanced strategies (for example betting different amounts based proportional to the odds) finding consistent results although increasing the risks and therefore the fluctuations in marginal profit.

%We now want to determine the performance of two systematic alternative betting strategies: the first baseline strategy is \emph{random betting}. This is a binary strategy that puts 50\% chance on each team, regardless of the odds. Is not a very realistic scenario, but serves as baseline. We randomly bet on the 56 games for 10 iterations. The results are shown in Fig. \ref{fig:profit2}: each bar shows the results based on random betting for each iteration.
%Although the profit can seldom be positive by chance, this strategy clearly yields big losses on the long term, with sometimes catastrophic performance (as bad as halving one's budget or worse).
%The average loss is $\mu=-9.2\%$ with a very large standard deviation of $\sigma=25.95\%$.

%\begin{figure}[!t]
%\centering
%\includegraphics[width=\columnwidth]{random.png}
%\caption{Profit/losses of random bets (10 rounds).} 
%\label{fig:profit2}
%\end{figure}

\begin{figure}[!t]
\centering
\includegraphics[width=\columnwidth]{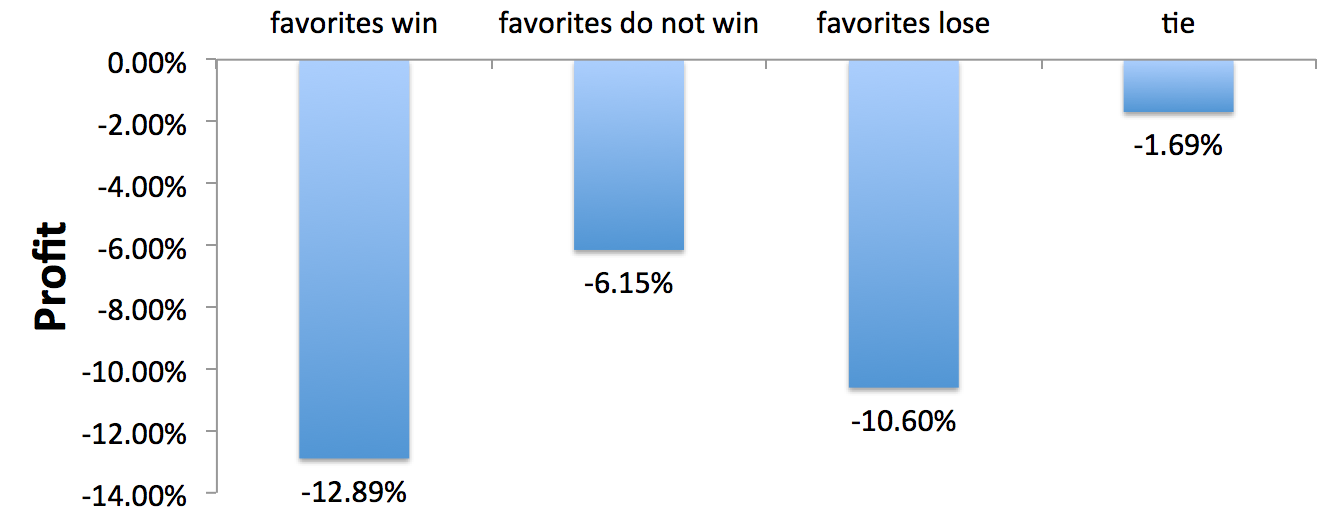}
\caption{Losses on bets using systematic strategies.}  
\label{fig:profit3}
\end{figure}

The final comparison is against systematic betting on the following results, independently from the game: \emph{(i)} the favorite team always wins, \emph{(ii)} the favorite team does not win (half dollar bet on a tie, half dollar bet on the underdog winning), \emph{(iii)} the favorite team loses (one dollar on the underdog winning), and \emph{(iv)} the match is a tie. % \red{, and finally,\emph{(v)} one of the two teams wins (half a dollar on the favorite and half a dollar on the underdog)}.
For each strategy we first compute the marginal profit on each game and then compute the average marginal profit (and the relative standard deviation) across all games.
The results are shown in Fig. \ref{fig:profit3}: these strategies, 
%although informed by the odds (and betting in line or against the odds) 
once again all yield possibly large losses. Interestingly enough, the safest fixed strategy (that still imposes a loss) is to bet on ties.

All benchmarks demonstrate that betting according to the predictions produced by our machine learning framework yields a consistent, positive and potentially large marginal profit, unparalleled by other systematic betting strategies, even informed by the odds.

\section{Data Collection} \label{sec:data}
We employed two different strategies for the collection of Twitter data relative to the 2014 FIFA World Cup and the other tournaments.
 
\subsection{Twitter data for the 2014 FIFA World Cup}
%To design our predictive framework based on social media streams, we had first to create a dataset to capture the conversations around the 2014 FIFA World Cup matches on Twitter.
World Cup games attracted much more global attention than any other soccer games before and after (indeed, of any other event ever, as previously noted) providing a very large data base.
We systematically collect and store all data from the Twitter \emph{gardenhose}, a 10\% sample of the entire Twitter stream. Focusing our search on the period  during which the World Cup occurred (June, 12th 2014 though July, 13th 2014), we isolated all tweets containing any of these keywords: \emph{(i)} the official abbreviation of the game, as recommended by FIFA\footnote{\url{http://www.fifa.com/worldcup/teams/index.html}}; \emph{(ii)} one or both of the team names; \emph{(iii)} one or both the official team abbreviations; or, \emph{(iv)} the hashtag combining the team names or abbreviation with ``vs'' (e.g., ``BRAvsGER'' to identify the game between Brazil and Germany). This procedure yielded a corpus of tweets for each of the 64 games occurred during the competition. 

\begin{figure}[!t]
\centering
\includegraphics[width=\columnwidth,clip=true,trim=30 30 30 30]{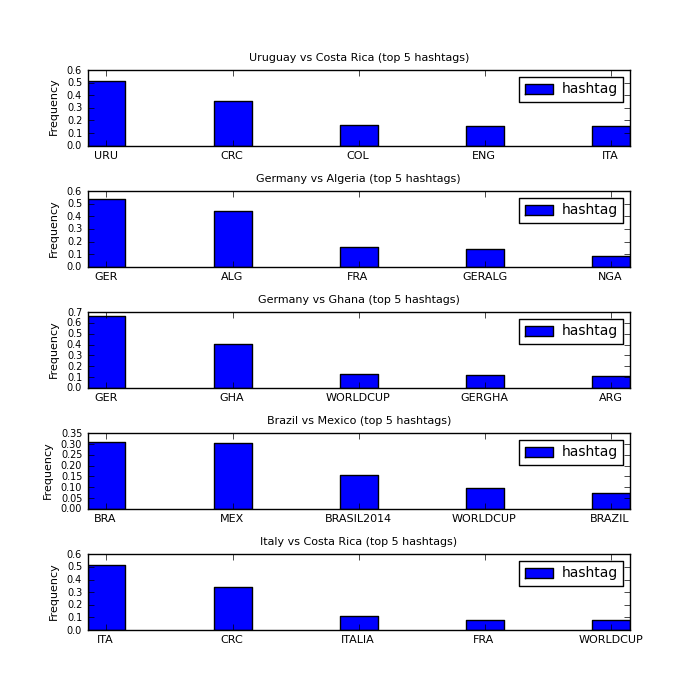}
\caption{Hashtags distribution for five upset games. The frequency is the percentage the tweets containing each hashtag among all collected tweets for that game. Co-occurrences yield sums larger than one.}
\label{fig:hashtags}
\end{figure}

%To understand the Twitter conversation that preceded World Cup matches, 
We isolated the tweets produced during the 6 hours before the beginning of each game and analyzed the frequency of adoption of the related keywords. The results for five representative matches are shown in Fig. \ref{fig:hashtags}. We noted that the abbreviations dominated the frequency of keywords adoption in all games. With a maximum limit of 140 characters per tweets, abbreviations are commonly used to save both space and typing time. Besides the team abbreviations, most of the other somehow frequent hashtags are either irrelevant (e.g., \#eng in the game of Uruguay vs. Costa Rica) or too general or broad (e.g., \#worldcup) to apply to the specific game itself. 
Therefore, we decided to use only hashtags of team abbreviations. Each game is therefore characterized by three subcategories of tweets: those related to each of the two teams involved in the match, and those related to the match itself (namely, those in which both team names appear).
We finally performed a manual validation of the dataset:  for all games, we randomly sampled 50 tweets in each of the three subsets and manually verified whether the tweets were correctly identified. In Table  \ref{tab:validation} we show the results of the validation procedure for five upset games.
%, where we also identify the favorite and the underdog team (based on the bookmakers' odds). 
Essentially all tweets collected with our procedure are closely related to the games. The precision is consistently above 90\% for every game in all the three subcategories. The final dataset of games for the 2014 FIFA World Cup contains 658,468 tweets, of which 319,312 are retweets and 28,707 are replies produced by 478,529 unique users.

\begin{table}[!t]
\centering \small
\caption{Manual validation of the quality of the 2014 FIFA World Cup dataset\bigskip}
\begin{tabular}{|c|r|r|r|}
\hline
Game	& Favorite	& Underdog	& Match\\
\hline\hline
\multirow{3}{*}{Uruguay vs Costa Rica}		&	94\% (Y) & 92\% (Y) & 100\%	(Y) \\
&	6\% (?) & 8\%  (?) & 0\%  (?) \\
&	0\% (N) & 0\%  (N) & 0\%  (N) \\
\hline
\multirow{3}{*}{Germany vs Algeria}		&	94\% (Y)  & 96\% (Y) & 100\%	 (Y) \\
&	6\%  (?) & 4\% (?)  & 0\%  (?) \\
&	0\% (N)  & 0\% (N)  & 0\% (N)  \\
\hline
\multirow{3}{*}{Germany vs Ghana}		&	96\%  (Y) & 96\% (Y) & 100\% (Y) \\
&	4\%  (?) & 4\% (?)  & 0\% (?)  \\
&	0\%  (N) & 0\% (N)  & 0\%  (N) \\
\hline
\multirow{3}{*}{Brazil vs Mexico}		&	92\% (Y)  & 94\% (Y) & 100\%	 (Y) \\
&	8\% (?)  & 6\%  (?) & 0\%  (?) \\
&	0\%  (N) & 0\% (N)  & 0\% (N)  \\
\hline
\multirow{3}{*}{Italy vs Costa Rica}		&	92\% (Y)  & 92\% (Y) & 100\%	 (Y) \\
&	8\% (?)  & 8\% (?)  & 0\%  (?) \\
&	0\% (N)  & 0\% (N)  & 0\% (N)  \\
\hline
\end{tabular}
\label{tab:validation}
\end{table}

%As a proof-of-concept of this idea, we designed a simple algorithm that monitors, in real time, the odds assigned to all the soccer games played in the four major European leagues: the English \emph{FA Premier League}, the Italian \emph{Serie A}, the Spanish \emph{La Liga}, and the German \emph{Bundesliga}. In addition, we monitor the 2014 UEFA Champions League games as well, as it sometimes happens that unbalanced games occur also in this tournament that attracts lot of attention. Once our system identifies a potential upset exceeding our threshold ($\theta=5$), it automatically starts tracking the Twitter conversation occurring around it. This process ends two hours after the game concludes: this allows us to collect data not only before the game, but also during and shortly after it. The details about the dataset collection are provided next.

\subsection{Live-monitoring soccer games data}
During the period between October, 25th 2014 and November, 26th 2014 we monitored the odds of all games for four European national tournaments (the English \emph{FA Premier League}, the Italian \emph{Serie A}, the Spanish \emph{La Liga}, and the German \emph{Bundesliga}), and the UEFA Champions League. Our system selected 55 potential upsets with a profitability of at least 5:1 ($\theta=5$) 
%in case the most unexpected result had occurred.
and we collected in real time tweets about these games using the Twitter Streaming API. 
%Among these games, 16 of them turned into upsets (the underdog either won or tied) while 39 were baseline games (the favorite won).
 As for the 2014 FIFA World Cup games, we selected  tweets based on hashtags containing teams' abbreviations. The adoption of the Streaming API ensured that we collected the entirety of relevant tweets, rather than a sample. Some games, however, do not have enough tweets to guarantee a meaningful analysis (for example because the involved teams are not very popular). We therefore filtered out those games that did not collect at least 40 tweets per team per hour. This post-processing yielded a dataset 31 games, of which 9 turned into upsets and 22 into baselines. The final dataset of league matches contains 1,278,485 tweets, including 521,776 retweets and 31,281 replies, produced by 1,009,034 unique users.

\subsection{Sentiment Analysis} \label{sub:sa}
The ability to capture and computationally represent supporters emotions and feelings, and how these evolve over time, is a crucial component of our system. In particular, the framework is designed to capture \emph{favorability} from content using sentiment analysis algorithms based on natural language processing~\cite{nasukawa2003sentiment} and opinion mining~\cite{pang2008opinion}. 
Previous studies have shown that sentiment analysis is able to capture the overall mood of a population and inform predictions about elections and financial markets movements~\cite{bermingham2011using,thelwall2011sentiment,bollen2011twitter}. 

After benchmarking the performance of the majority of sentiment analysis libraries available, we determined that the most suitable for our system is the \emph{Indico deep learning} sentiment analysis framework, and we adopted the relative Python API\footnote{\url{https://pypi.python.org/pypi/IndicoIo/0.4.7}}.
The algorithm returns a sentiment score between 0 and 1 for each tweet. 
We evaluated its performance using the \emph{Stanford Twitter sentiment corpus} (STS-test)\footnote{\url{http://help.sentiment140.com/for-students}}, a manually annotated dataset containing 177 negative, 182 positive and 139 neutral tweets~\cite{saif2013evaluation}. 
The STS-test is relatively small but it has been widely used to benchmark several sentiment analysis algorithms~\cite{saif2013evaluation,saif2011semantic,saif2012alleviating,go2009twitter,speriosu2011twitter,bakliwal2012mining}.  As a comparison example we report the performance of \emph{Text-Processing}\footnote{\url{http://text-processing.com/docs/sentiment.html}}, a sentiment tool trained on both Twitter data and movie reviews\footnote{\url{http://www.cs.cornell.edu/people/pabo/movie-review-data/}} adopting a Naive Bayes classifier. 
The results of the benchmarks are shown in Table \ref{tab:sentiment}. 
Indico outperforms Text-Processing (and all other algorithms we tested) achieving above 80.5\% accuracy, the highest ever reported on the STS-test~\cite{saif2013evaluation}, when we label as neutral all tweets with sentiment score comprised between 0.3 and 0.7.
Hereafter, we use this configuration.

\begin{table}[!t]
\centering\small
\caption{Sentiment tools performance on STS-test\bigskip}
\begin{tabular}{|c|c|c|}
\hline
Algorithm	&	Accuracy	&	Configuration\\
\hline\hline
Text-Processing & 0.6045 & no neutral tweets \\
\hline
Indico  & 0.7465 & no neutral tweets\\
\hline
Indico  & 0.7088 & neutral: between 0.4 and 0.6 \\
%\hline
%Text-Processing & 0.6124 & /\\
\hline\rowcolor{Gray}
Indico  & 0.8052 & neutral: between 0.3 and 0.7\\
\hline
\end{tabular}
\label{tab:sentiment}
\end{table}

\section{Related Work} \label{sec:related}
This work, to the best of our knowledge, is the first to exploit social media streams to predict soccer matches. However, various recent studies have approached related problems~\cite{tumasjan2010predicting}, such as predicting the outcome of political elections~\cite{digrazia2013more,schoen2013power}, talent shows~\cite{ciulla2012beating}, movies success~\cite{asur2010predicting,oghina2012predicting}, stock-market fluctuations~\cite{zhang2011predicting,bollen2011twitter}, political protests~\cite{choudhary2012social,conover2013geospatial,conover2013digital,varol2014evolution}, and diffusion of information~\cite{mathioudakis2010twittermonitor,ferrara2013traveling}. 

To prove the idea that social media data convey predictive power, Asur and Huberman~\cite{asur2010predicting} designed a system that uses Twitter to forecast the box-office revenues of upcoming movies: simple signals such as the buzz around a given movie seem indicative of its future popularity. DiGrazia {\em{et al.}}~\cite{digrazia2013more} used a similar framework to show that there exists a statistically significant association between tweets that mention a political candidate for the U.S. House of Representatives and his or her subsequent electoral performance. Bermingham and Smeaton~\cite{bermingham2011using} illustrated a similar case study for the recent Irish General Election, modeling political sentiment by mining social media conversations. They combined sentiment analysis using supervised learning and volume-based measures and found that this signals are highly predictive of election results. Bollen {\em{et al.}}~\cite{bollen2011twitter} analyzed the textual content of the daily Twitter stream to show that Twitter mood is predictive of the daily fluctuations in the closing values of the Dow Jones Industrial Average (DJIA). Xue Zhang {\em{et al.}}~\cite{zhang2011predicting} collected Twitter data for six months and found that the percentage of emotional tweets significantly negatively correlates with Dow Jones, NASDAQ and S\&P 500 fluctuations, but displays a significant positive correlation to VIX.

Various works called for caution when using social media to predict exogenous events~\cite{gayo2012no,gayo2011limits}: in such cases, it is important to keep in mind that the usage of machine learning algorithms or statistical models that function as black boxes can yield to results which are not interpretable and misleading~\cite{lazer2014parable}. 
For these reasons, when we designed our machine learning framework we based it on simple assumptions: the prediction dynamics are entirely explainable and observable in real time. In fact, our model relies only on one single feature (the average conversation sentiment measured over time) and it allows to interpret the predictions in a concise and clear way. Our hypotheses are also rooted on recent advances in social psychology that support the idea that collective attention enhances group emotions~\cite{mason2007situating,shteynberg2013power,shteynberg2014feeling}.
 
\section{Conclusions}
In this paper we presented a machine learning framework that leverages social media signal to effectively predict the outcome of very unbalanced games.
 
We analyzed Twitter conversations relative to potential upset games to provide evidence that signal extracted from the conversation reflects the sentiment of the large crowd of fan following the game.  
%We presented two scenarios: our first analysis was carried out using historical Twitter data about games occurred during the 2014 FIFA World Cup. We then designed a system to track the four major European tournaments (English FA Premier League, Italian Serie A, Spanish La Liga, and German Bundesliga) and the UEFA Champions League, live-monitoring all potential upsets for one month (October, 25th 2014 to November, 26th 2014), and collecting in real-time the social media conversation about the matches.
We showed that our systems achieves a very promising prediction performance, with accuracy and AUROC around 80\%.
We also demonstrated that the predictions yielded by our system
%only using Twitter signals up until 30 minutes before the games start, 
can be effectively used to inform betting strategies achieving a positive and not negligible profit above 8\%, and compared it with a number of baseline strategies that invariably leads to losses. We deem this as a strict and rigorous test of the effectiveness of our method.

Beating the odds offered by bookmakers is notoriously difficult, and is certainly not by chance that the betting industry is large and very profitable.
Professional bookmakers matured great expertise in setting the initial odds, can readjust quotes continuously according to the incoming bets, and grant themselves a generous profit margin. We believe that the reason for our success relies, in part, in focusing on very unbalanced games, where at least one of the potential results is deemed as highly unlikely. The high unlikelihood of one the result may lead to an increased difficulty in correctly estimating the relative odd. Also the exploitability of very unbalanced games could be the consequence of a general aversion in the betting crowd towards betting on unlikely results: this would lead to enhanced odds for the unlikely result to attract bets that can offset the losses incurred by the bookmaker if the most likely (and most bet upon) result materializes. We may imagine that profit margin would decrease if we apply our method to a set of more balanced games, and plan to test our hypothesis extensively in future work.

\section*{Acknowledgments}
AF acknowledges support by NSF Award No. IIS-0811994.

%\balancecolumns
\bibliographystyle{abbrv}
\bibliography{bib} 
\end{document}